\documentclass[a4paper,11pt]{article}
\usepackage{pos}

\usepackage{xcolor}

\title{Transient Observations with LST-1: Key Results and Future Prospects}
\manuallySeparateAuthors

\author*[a]{Monica Seglar-Arroyo}
\author[b]{Alessio Berti,}
\author[c]{Alessandro Carosi,}
\author[d]{Gloria Maria Cicciari,}
\author[c]{Alice Donini,}
\author[e]{Armand Fiasson,}
\author[f]{Arnau Aguasca-Cabot,}
\author[g]{Mathieu de Bony de Lavergne,}
\author[f]{Pol Bordas,}
\author[h]{Alicia Lopez-Oramas,}
\author[f]{Marc Ribó,}
\author[e]{Edna Ruiz-Velasco,}
\author[i]{Fabian Sch\"ussler,}
\author[]{on behalf of the CTAO-LST Collaboration}
\affiliation[a]{Institut de Fisica d’Altes Energies (IFAE), BIST, Bellaterra (Barcelona), Spain}
\affiliation[b]{Max-Planck-Institut für Physik, Garching bei München}
\affiliation[c]{INAF - Osservatorio Astronomico di Roma, Monteporzio Catone, Italy}
\affiliation[d]{Dipartimento di Fisica e Chimica 'E. Segrè' Università degli Studi di Palermo, Palermo}
\affiliation[e]{Laboratoire d’Annecy de Physique des Particules (LAPP), U. Savoie Mont Blanc, CNRS/IN2P3, France.}
\affiliation[f]{Departament de Física Quàntica i Astrofísica, Institut de Ciències del Cosmos, Universitat de Barcelona, IEEC-UB, Barcelona, Spain.}
\affiliation[g]{Centre de Physique des Particules de Marseille (CPPM), Aix–Marseille Université, CNRS/IN2P3, 163, Marseille, France}
\affiliation[h]{{Instituto de Astrofísica de Canarias and Departamento de Astrofísica, Universidad de La Laguna, C. Vía Láctea, s/n, 38205 La Laguna, Santa Cruz de Tenerife, Spain}}
\affiliation[i]{IRFU, CEA, Université Paris-Saclay, Gif-sur-Yvette, France}

\emailAdd{mseglar@ifae.es}

\abstract{The recent detections of the afterglow phase of long gamma-ray bursts at very high energies (VHE, >100 GeV) mark a significant advance in astrophysics of transient phenomena, offering deeper insights into the acceleration mechanisms, jet structure, and physical processes driving GRB emission. In the multi-messenger landscape, both high-energy neutrino and gravitational wave detections are providing new insights into the physics of extreme cosmic accelerators and highlighting the need for rapid and broadband follow-up observations. The Large-Sized Telescope (LST-1), the first telescope of the LST array, part of the Cherenkov Telescope Array Observatory (CTAO) North site, is particularly well-suited for real-time, rapid follow-up of transients.  In this contribution, we present the latest achievements of the transient observational program with LST-1, which is now in advanced commissioning on La Palma, Canary Islands. We outline the observational strategies in place and describe the dynamic handling of events by the transient handler of LST-1 (e.g., its ability to handle poorly localised events, including gravitational waves, GRBs and neutrinos). We present the key results from transient observation campaigns conducted so far, discuss the lessons learned, and outline the promising prospects for the future LST-1+MAGIC combined transient program with fast response, via a Transient Handler.}

\ConferenceLogo{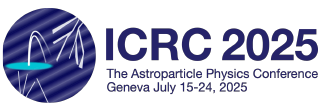}

\FullConference{39th International Cosmic Ray Conference (ICRC2025)\\
 15–24 July 2025\\
Geneva, Switzerland\\}

\begin{document}
\maketitle
\section{Introduction}

Transient phenomena in the Universe—brief, energetic events that evolve on timescales ranging from milliseconds to days—offer unique insights into the most extreme physical processes. Sources exhibiting this behaviour in the gamma-ray regime include gamma-ray bursts (GRBs), as probed by the recent GRB detections at VHEs—GRB 180720B \cite{GRB180720B}, GRB 190114C \cite{GRB190114C}, GRB 190829A \cite{GRB190829A}, and GRB 201216C \cite{GRB201216C}. This progress reached a new peak with the detection of GRB 221009A \cite{GRB221009A}, the brightest GRB ever observed. While GRBs remain a flagship target, the landscape of transient astrophysics has expanded in recent years. The advent of multi-messenger astronomy has revealed a broader population of energetic transients that are prime candidates to emit at gamma-ray energies (GeV--TeV). The discovery of gravitational waves (GWs) in 2015 from compact binary mergers GW150914 \cite{GW150914} by LIGO and Virgo— enabled the link of short GRBs to neutron star mergers, with the GW170817–GRB 170817A association \cite{GWGRB2017}. In parallel, high-energy neutrino alerts from observatories like IceCube have become increasingly important triggers for rapid follow-up in the gamma-ray domain. Events such as IceCube-170922A, potentially associated with the blazar TXS0506+056 \cite{TXS}, or the tidal disruption event (TDE) created by a black hole tearing a star apart, AT2019dsg as identified by the Zwicky
Transient Facility, associated with the neutrino IC 191001A \cite{TDE}, underscore the importance and potential of chasing transient neutrino events.  Moreover, fast radio bursts (FRBs), and Galactic transients such as core-collapse supernovae (CCSNe), are other targets of interest in transient observation campaigns in gamma rays. Concerning fast-evolving galactic transients, novae have been established as gamma-rays emitters, both from \textit{Fermi}-LAT GeV detection of tens of binary systems, including classical and symbiotic novae, but specially at VHEs after the symbiotic nova RS Oph detection in 2021 by MAGIC \cite{MAGICRS}, H.E.S.S.\cite{HESSRS} and LST-1 \cite{LST1RS}.  

These diverse classes of transients pose both opportunities and challenges for ground-based gamma-ray observatories. 
Dedicated frameworks and software—such as tiling algorithms, real-time data analysis pipelines, and alert-driven scheduling—have become essential components of modern VHE observational systems. In this evolving landscape, the first telescope of the four Large-Sized Telescopes (LSTs) of the Cherenkov Telescope Array Observatory in the North (CTAO-North), LST-1, is producing scientific results and aims to enhance its capabilities to respond to a wide array of transient multi-messenger triggers. These efforts are crucial for probing the most extreme physical processes in the universe, from black hole and neutron star mergers to relativistic jets and shock acceleration in dense environments.

\section{The first Large-Sized Telescope (LST-1)}

The CTAO, the leading gamma-ray Cherenkov telescope observatory of the next decades, will consist of two multi-telescope arrays placed in two distinct sites—one in the North (La Palma, Spain) and another in the South (Atacama, Chile)—covering a very large portion of the sky and allowing almost continuous operations. Its unprecedented performance comes from its multi-telescope design, based on a combination of LSTs, Medium-Sized Telescopes (MSTs) and Small-Sized Telescopes (SSTs). These cover the energy range in ascending order. The LST design makes it particularly well suited to cover transient, extragalactic sources thanks to their improved sensitivities at short time scales compared to space-based instruments, to its large reflective surface, enabling the detection of faint Cherenkov flashes from electromagnetic cascades of gamma rays down to 20 GeV, its 4.5 deg field-of-view camera, allowing coverage of broad sky regions, and the light design of the telescope structure, allowing for fast slewing between coordinates, up to 180 deg in about 20 s. 

\subsection{The transient program of LST-1.} The current LST-1 transient program searches for gamma-ray emission from various distinct sources, in order of decreasing reaction time. After the confirmation of the long-standing hypothesis that some GRBs are VHE emitters, we aim to broaden this knowledge by leveraging the capabilities of LST-1. We target alerts from various instruments, in decreasing energy range coverage: LHAASO, HAWC, \textit{Fermi}-LAT, \textit{SVOM}-Eclair, \textit{Fermi}-GBM, \textit{SVOM}-GRM, \textit{Swift}-BAT, \textit{Swift}-XRT, MAXI, and Einstein Probe, as well as optical telescopes information, for magnitude and redshift estimation. For GW, our follow-up program covers all the alert types provided by the LIGO-Virgo-KAGRA (LVK) Collaboration, including compact object in binaries, as binary neutron stars (BNS), neutron start black hole mergers (NSBH) and binary black hole mergers (BBH), including those in the sub-solar mass range and the mass gap, and bursts.  Regarding the neutrino program, both the publicly distributed neutrino, via General Coordinates Network (GCN)/AMON (neutrino tracks and neutrino cascades)\cite{AMON}, as well as neutrinos privately shared via Memorandum of Understanding (MoU), including pre-selected sources and all-sky neutrino flares, are considered. In addition, we follow up significant AMON alerts, with a special focus on Icecube$+$HAWC alerts. Building on recent TDE-neutrino associations, we aim to investigate the scenarios where TDEs might produce gamma rays,  which requires particle acceleration. Accordingly, we focus on jetted TDEs, TDEs associated with neutrino counterparts, and also opportunistic nearby TDEs within $\sim$100 Mpc. Regarding FRBs, we target FRB repeaters during active periods to ensure the widest possible multi-wavelength coverage. Lastly, the core-collapse supernova (CCSN) program covers the most promising CCSN types that could potentially power gamma-ray emission, particularly at times when the $\gamma\gamma$ attenuation decreases and the gamma-ray emission is expected to reach its maximum value. The most promising CCSNe are those similar to SN 1987A. The Transient Name Server (TNS) \cite{TNS} is used to identify them, using a newly developed filtering pipeline. In the case of novae, the Central Bureau Astronomical Telegrams’ Transient Objects Confirmation Page, VSNET \cite{VSNET}, and AAVSO \cite{AAVSO} notices together with reports of transients through The Astronomer's Telegram (ATel) \cite{ATel} and TNS are monitored to identify detection opportunities indicated by a flux increase, with particular attention to \textit{Fermi}-LAT.

The LST-1 transient program is supported by the LST-1 Transient Handler (TH), which serves as the core system for follow-ups—coordinating the various subsystems involved in the transient response and enabling automatic, real-time observation scheduling. This is complemented by the offline human-in-the-loop response—assured by a team of experts—, whenever the alert arrives during the day or in scientific cases where mid-cadences are the most adapted strategy. Regarding source localisations, well-localised transients, such as \textit{Fermi}-LAT or \textit{Swift}-BAT GRBs or IceCube neutrino tracks, the response strategy is straightforward: the telescope repoints to the target, and the observation is conducted using standard wobble mode (e.g., 20-minute runs with an offset pointing to allow simultaneous background estimation). In contrast, poorly localised events—typical of \textit{Fermi}-GBM GRBs, neutrino cascades, or most GW alerts—require more advanced pointing strategies. In such cases, sky tiling becomes essential, for which the Python package \texttt{tilepy} \cite{tilepy} is used. The offline scheduling is streamlined and automated through the use of Astro-COLIBRI \cite{AC}, which acts as a bridge, automatically populating the observation request to be sent to schedulers with all relevant information extracted from the alert.

\subsection{The LST-1 Transient Handler.} 
LST-1 TH has the primary role of receiving, processing and initiating an automatic reaction of the telescope that enables the observations of transient sources. The system has been handling transient alerts since $\sim$2020 \cite{Carosi2021}. It is continually evolving in terms of new protocols and alert types. Currently, alerts are provided by GCN via the event streaming platform Kafka \cite{kafka}, supporting both VOEvents (XML) and JSON formats. Source-adapted selection cuts are applied on different parameters. These are provided by a team of experts leading the observation proposal. These consist of timing, visibility of the source, and parameters of the detected source included in the alert, as the false alarm rate (FAR) or the probability that the source originates from a given astrophysical population. The TH workflow has been designed to react fully automatically to the various types of alert and their updates. A colour coding in the schematic diagram in Figure~\ref{fig:LSTTH} is used to differentiate the main components and data flow within the transient observation framework:

\begin{figure}[h!]
\centering
\includegraphics[width=0.8\columnwidth]{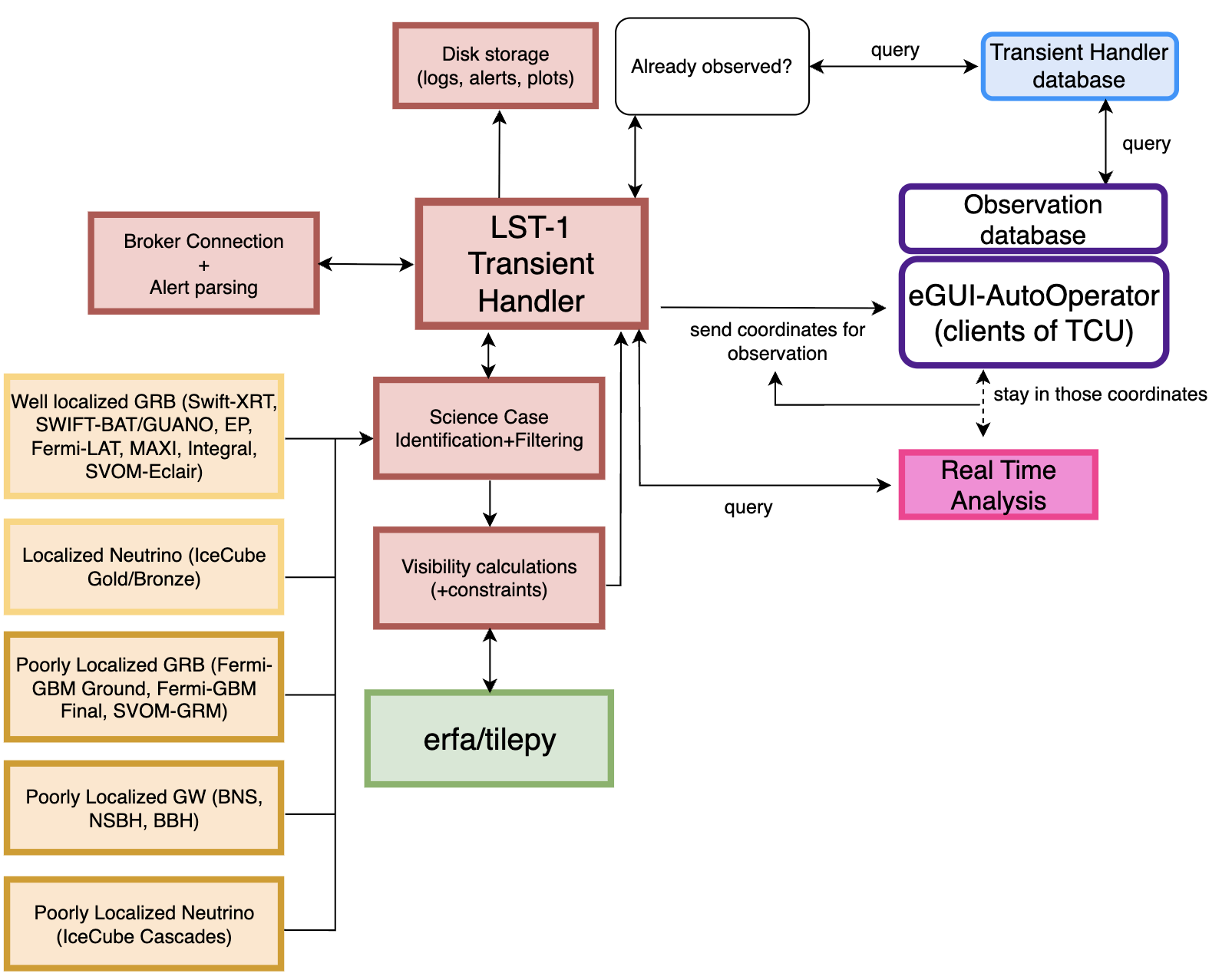}
\caption{Workflow of the LST-1 TH. The different colours correspond to the various connected systems that give an effective, dynamic transient response (see text).}
\label{fig:LSTTH}
\end{figure}

\begin{itemize}
    \item \textbf{LST-1 Transient Handler (red):} The core functionalities include the alert reception, the prioritisation, and the scheduling management, forming the central decision-making engine of the system. The science case identification and filtering are based on the criteria in the scientific proposals submitted for each observing cycle. Visibility calculations are also handled so moonlight and darktime observations are identified.
    
    \item \textbf{Config files (yellow):} These include the parameters required to define the reaction behavior for each science case enabling tailored observation strategies based on the alert content.

    \item \textbf{\texttt{tilepy} (\cite{tilepy}, green):} The open-source Python package is used as the sky-tiling engine to compute optimised pointing patterns, used for sources with large localisation uncertainties.

    \item \textbf{Telescope Control Unit (TCU, \cite{TCU}, purple):} This interface is responsible for commanding the various telescope systems and launching observations. Hence, it allocates the scheduled pointings accordingly. The TH sends the required information to the system managing telescope observations, known as AutoOperator (AO). The AO creates a corresponding source configuration and observation block, and executes it at the requested time.
    
    \item \textbf{Transient Handler Database (blue):} This component allows the system to retrieve and track previously scheduled observations by the TCU. It supports the handling of alert updates, based on the ongoing or past observations associated with the same event.
    
    \item \textbf{Real-Time Analysis (RTA, \cite{RTA}, pink):} The system is in charge of the real-time data analysis, which processes incoming data on short timescales to evaluate the presence of gamma-ray emission detection during ongoing observations, providing rapid feedback to the system.
\end{itemize}

\begin{figure}[h!]
\centering
\includegraphics[width=0.55\columnwidth]{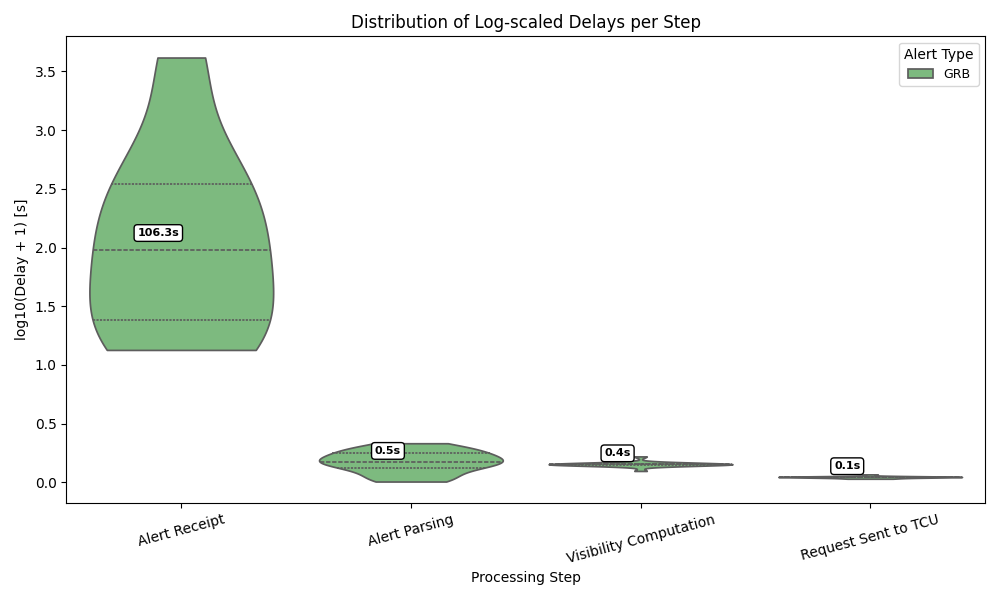}
\includegraphics[width=0.44\columnwidth]{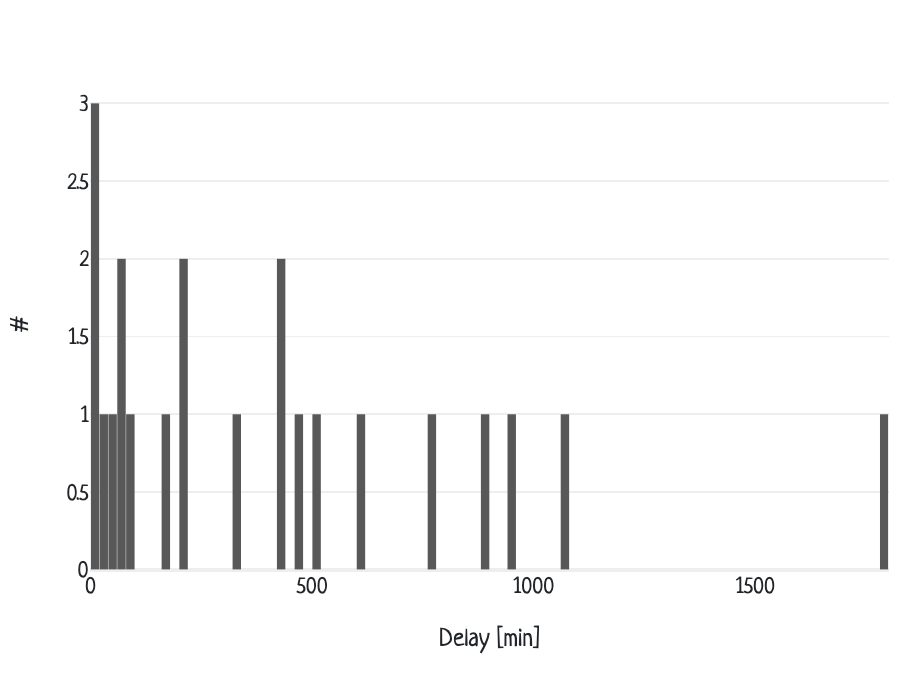}
\caption{Left: TH reaction times per step.  Right: Reaction time to the scheduled observations/to change by TH reaction times).}
\label{fig:Statistics}
\end{figure}

\section{LST-1 observations}

Since the end of 2023, a total of 22 alerts have been followed by LST-1.  GRBs are the largest of these, with 20 observed alerts ($\sim$ 48 h) and two GW sources ($\sim$ 3.3 h). These observations were only marginally affected by telescope-related issues, as the commissioning phase had significantly progressed. The sky localisation distribution of these events is shown in Figure \ref{fig:Events_S240615dg}. Out of these, we highlight the observations on the exceptionally bright, long GRB 2210009A, detected by \textit{Fermi}-GBM, \textit{Swift}-BAT and at VHE gamma rays by LHAASO. It was observed by LST-1 for 3.17\,h under bright moonlight conditions at T$_0$+1.33 days, and monitored until the end of November 2022, due to the exceptional nature and brightness of the GRB. The analysis results can be found in \cite{GRB221009A}. Secondly, we highlight the LST-1 observations, coordinated with the MAGIC telescopes, of two BBH candidates detected by LVK during the O4 run S240615dg \cite{S240615dg} and S241125n \cite{S241125n}, $\sim$ 14 hours and $\sim$ 19 hours after the initial trigger, respectively. S240615dg is the best localised event of O4 (50$\%$ C.R. of $\sim$1 deg$^2$), while S241125n (50$\%$ C.R. of >1 deg$^2$) is the only event in O4 with a candidate counterpart identified, in X-rays, by the \textit{Swift}-BAT~\citep{S241125nS}. An upcoming publication will report the results of these observations. 
 
\begin{figure}[h!]
\centering
\includegraphics[width=0.42\columnwidth]{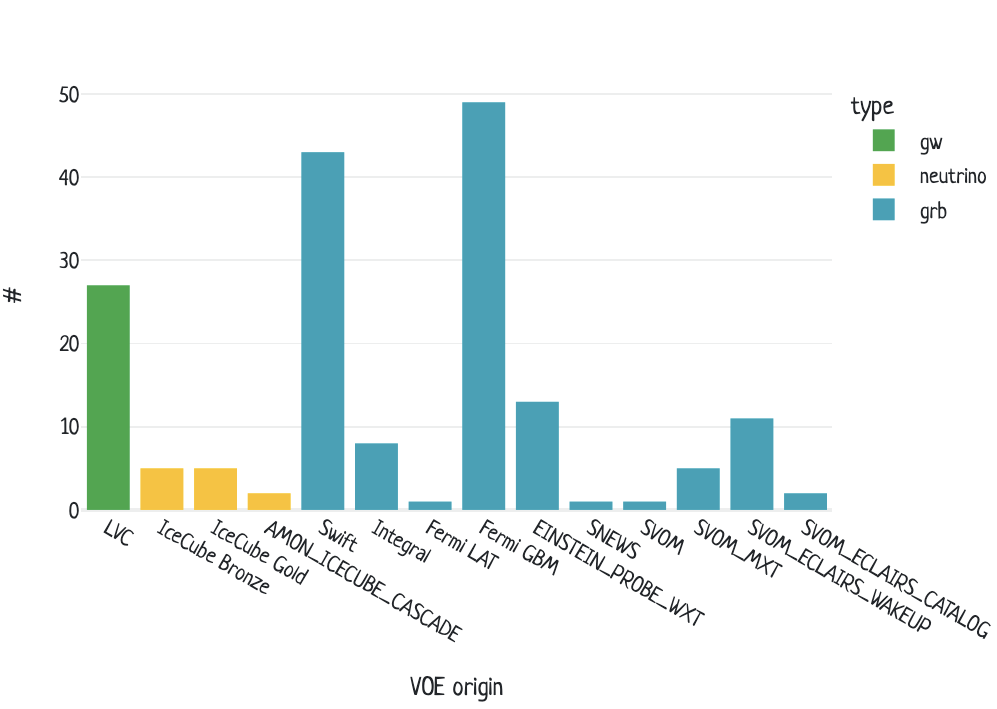}
\includegraphics[width=0.57\columnwidth]{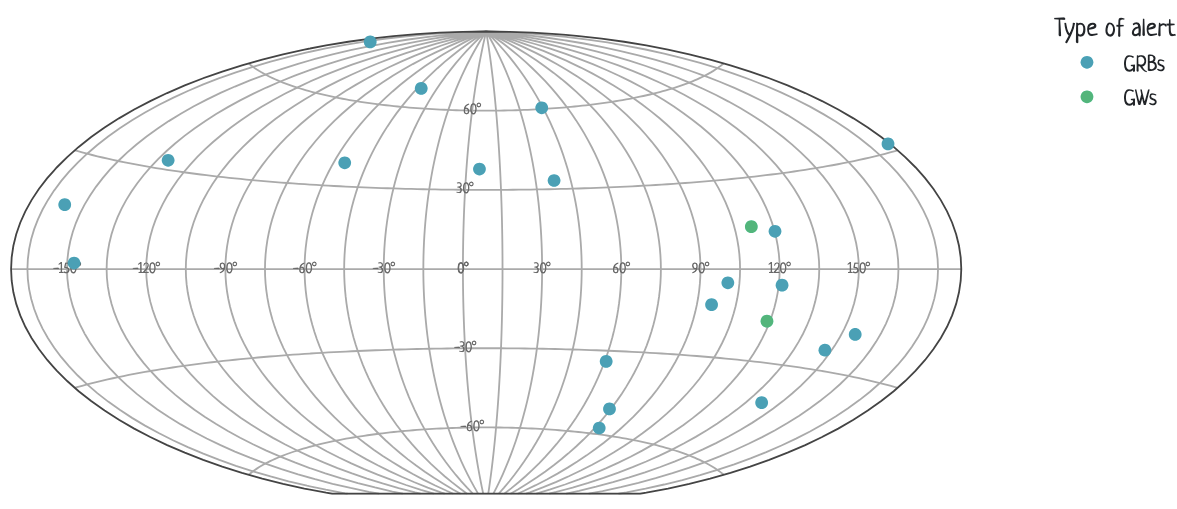}
\caption{Left: Distribution according to the VoEvent ID of the 173 alerts tagged as \texttt{OBSERVABLE} since December 2023. Right: Sky location of the GRBs and GW in Galactic coordinates (in the case of poorly localised events, only the largest probability pixel is represented).}
\label{fig:Events_S240615dg}
\end{figure}

\section{Joint LST-1 follow-up campaigns with MAGIC} 
The MAGIC and CTAO-LST Collaborations worked towards a joint observing cycle starting in April 2025. Joint transient observations required developments at all levels: from the target of opportunity infrastructure and shared communication platforms, to the definition of common tools, to facilitate rapid response and reduce duplication of effort. The automatic response has been designed to have the LST-1 TH as the orchestrator. The automatic reaction in well-localised targets is performed following the usual method, where the directions of the wobbled observations are selected beforehand for consistency. Regarding poorly localised transient events, these are covered by \texttt{tilepy}. The package is employed to compute efficient tiling patterns and coordinate multi-telescope scheduling when needed. An standout example of this collaborative approach is the wobble-mode GW tiling campaign on S240615dg with an offline-coordinated tiling pattern and synchronised scheduling, to cover over 95\% of the GW localisation region. The capabilities of this coordinated, complementary LST-1 with MAGIC tiling are depicted using the case of GRB 240612A, for which we could achieve a coverage of 83\% of the source location uncertainty region, paving the way of GRB detection despite challenging localisation. These are shown in Figure \ref{fig:Events_S240615dg}. 

\begin{figure}[h!]
\centering
\includegraphics[width=0.33\columnwidth]{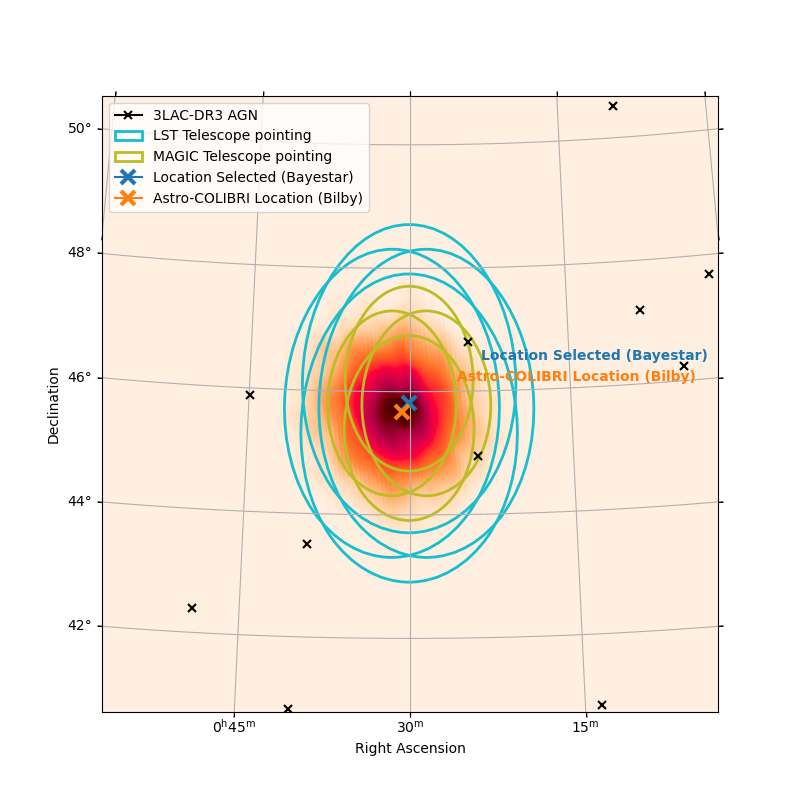}
\includegraphics[width=0.66\columnwidth]{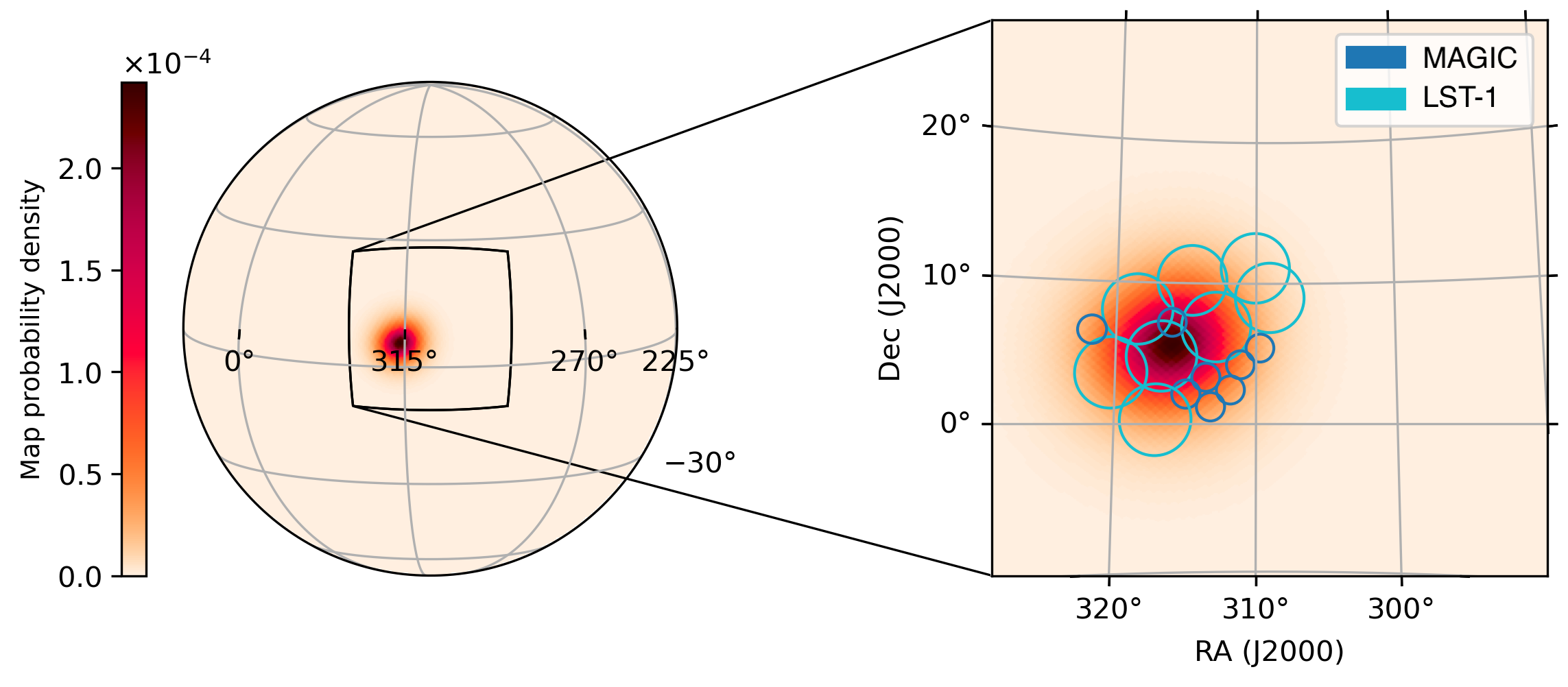}
\caption{Left: Joint wobble LST-1 and MAGIC observations of the GW S240615dg. Right: Example of a joint LST-1 and MAGIC follow-up campaign on GRB 240612A (\texttt{bn240612503}).}
\label{fig:Events_S240615dg}
\end{figure}

\section{Outlook} Aimed at shedding light on the underlying astrophysical mechanisms in transient sources, the LST-1 Transient program has reached a mature state with 9 scientific programs in conjunction with the MAGIC Collaboration. The remaining three LST telescopes designed for the CTAO-North site are under construction, with the first stereo LST data confidently expected by the end of 2026.

\vspace{5mm} 

\textbf{Full Author List: CTAO-LST Project}
\tiny{\noindent
K.~Abe$^{1}$,
S.~Abe$^{2}$,
A.~Abhishek$^{3}$,
F.~Acero$^{4,5}$,
A.~Aguasca-Cabot$^{6}$,
I.~Agudo$^{7}$,
C.~Alispach$^{8}$,
D.~Ambrosino$^{9}$,
F.~Ambrosino$^{10}$,
L.~A.~Antonelli$^{10}$
% ...
The CTAO-LST Project
}\\

\tiny{\noindent$^{1}${Department of Physics, Tokai University, 4-1-1, Kita-Kaname, Hiratsuka, Kanagawa 259-1292, Japan}.
$^{2}${Institute for Cosmic Ray Research, University of Tokyo, 5-1-5, Kashiwa-no-ha, Kashiwa, Chiba 277-8582, Japan}.
$^{3}${INFN and Università degli Studi di Siena, Dipartimento di Scienze Fisiche, della Terra e dell'Ambiente (DSFTA), Sezione di Fisica, Via Roma 56, 53100 Siena, Italy}.
$^{4}${Université Paris-Saclay, Université Paris Cité, CEA, CNRS, AIM, F-91191 Gif-sur-Yvette Cedex, France}.
$^{5}${FSLAC IRL 2009, CNRS/IAC, La Laguna, Tenerife, Spain}.
$^{6}${Departament de Física Quàntica i Astrofísica, Institut de Ciències del Cosmos, Universitat de Barcelona, IEEC-UB, Martí i Franquès, 1, 08028, Barcelona, Spain}.
$^{7}${Instituto de Astrofísica de Andalucía-CSIC, Glorieta de la Astronomía s/n, 18008, Granada, Spain}.
$^{8}${Department of Astronomy, University of Geneva, Chemin d'Ecogia 16, CH-1290 Versoix, Switzerland}.
$^{9}${INFN Sezione di Napoli, Via Cintia, ed. G, 80126 Napoli, Italy}.
$^{10}${INAF - Osservatorio Astronomico di Roma, Via di Frascati 33, 00040, Monteporzio Catone, Italy}.
% ...
}

\acknowledgments 
\tiny{
We gratefully acknowledge financial support from the following agencies and organisations:
Conselho Nacional de Desenvolvimento Cient\'{\i}fico e Tecnol\'{o}gico (CNPq), Funda\c{c}\~{a}o de Amparo \`{a} Pesquisa do Estado do Rio de Janeiro (FAPERJ), Funda\c{c}\~{a}o de Amparo \`{a} Pesquisa do Estado de S\~{a}o Paulo (FAPESP), Funda\c{c}\~{a}o de Apoio \`{a} Ci\^encia, Tecnologia e Inova\c{c}\~{a}o do Paran\'a - Funda\c{c}\~{a}o Arauc\'aria, Ministry of Science, Technology, Innovations and Communications (MCTIC), Brasil;
Ministry of Education and Science, National RI Roadmap Project DO1-153/28.08.2018, Bulgaria;
Croatian Science Foundation (HrZZ) Project IP-2022-10-4595, Rudjer Boskovic Institute, University of Osijek, University of Rijeka, University of Split, Faculty of Electrical Engineering, Mechanical Engineering and Naval Architecture, University of Zagreb, Faculty of Electrical Engineering and Computing, Croatia;
Ministry of Education, Youth and Sports, MEYS  LM2023047, EU/MEYS CZ.02.1.01/0.0/0.0/16\_013/0001403, CZ.02.1.01/0.0/0.0/18\_046/0016007, CZ.02.1.01/0.0/0.0/16\_019/0000754, CZ.02.01.01/00/22\_008/0004632 and CZ.02.01.01/00/23\_015/0008197 Czech Republic;
CNRS-IN2P3, the French Programme d’investissements d’avenir and the Enigmass Labex, 
This work has been done thanks to the facilities offered by the Univ. Savoie Mont Blanc - CNRS/IN2P3 MUST computing center, France;
Max Planck Society, German Bundesministerium f{\"u}r Bildung und Forschung (Verbundforschung / ErUM), Deutsche Forschungsgemeinschaft (SFBs 876 and 1491), Germany;
Istituto Nazionale di Astrofisica (INAF), Istituto Nazionale di Fisica Nucleare (INFN), Italian Ministry for University and Research (MUR), and the financial support from the European Union -- Next Generation EU under the project IR0000012 - CTA+ (CUP C53C22000430006), announcement N.3264 on 28/12/2021: ``Rafforzamento e creazione di IR nell’ambito del Piano Nazionale di Ripresa e Resilienza (PNRR)'';
ICRR, University of Tokyo, JSPS, MEXT, Japan;
JST SPRING - JPMJSP2108;
Narodowe Centrum Nauki, grant number 2023/50/A/ST9/00254, Poland;
The Spanish groups acknowledge the Spanish Ministry of Science and Innovation and the Spanish Research State Agency (AEI) through the government budget lines
PGE2022/28.06.000X.711.04,
28.06.000X.411.01 and 28.06.000X.711.04 of PGE 2023, 2024 and 2025,
and grants PID2019-104114RB-C31,  PID2019-107847RB-C44, PID2019-104114RB-C32, PID2019-105510GB-C31, PID2019-104114RB-C33, PID2019-107847RB-C43, PID2019-107847RB-C42, PID2019-107988GB-C22, PID2021-124581OB-I00, PID2021-125331NB-I00, PID2022-136828NB-C41, PID2022-137810NB-C22, PID2022-138172NB-C41, PID2022-138172NB-C42, PID2022-138172NB-C43, PID2022-139117NB-C41, PID2022-139117NB-C42, PID2022-139117NB-C43, PID2022-139117NB-C44, PID2022-136828NB-C42, PDC2023-145839-I00 funded by the Spanish MCIN/AEI/10.13039/501100011033 and “and by ERDF/EU and NextGenerationEU PRTR; the "Centro de Excelencia Severo Ochoa" program through grants no. CEX2019-000920-S, CEX2020-001007-S, CEX2021-001131-S; the "Unidad de Excelencia Mar\'ia de Maeztu" program through grants no. CEX2019-000918-M, CEX2020-001058-M; the "Ram\'on y Cajal" program through grants RYC2021-032991-I  funded by MICIN/AEI/10.13039/501100011033 and the European Union “NextGenerationEU”/PRTR and RYC2020-028639-I; the "Juan de la Cierva-Incorporaci\'on" program through grant no. IJC2019-040315-I and "Juan de la Cierva-formaci\'on"' through grant JDC2022-049705-I. They also acknowledge the "Atracci\'on de Talento" program of Comunidad de Madrid through grant no. 2019-T2/TIC-12900; the project "Tecnolog\'ias avanzadas para la exploraci\'on del universo y sus componentes" (PR47/21 TAU), funded by Comunidad de Madrid, by the Recovery, Transformation and Resilience Plan from the Spanish State, and by NextGenerationEU from the European Union through the Recovery and Resilience Facility; “MAD4SPACE: Desarrollo de tecnolog\'ias habilitadoras para estudios del espacio en la Comunidad de Madrid" (TEC-2024/TEC-182) project funded by Comunidad de Madrid; the La Caixa Banking Foundation, grant no. LCF/BQ/PI21/11830030; Junta de Andaluc\'ia under Plan Complementario de I+D+I (Ref. AST22\_0001) and Plan Andaluz de Investigaci\'on, Desarrollo e Innovaci\'on as research group FQM-322; Project ref. AST22\_00001\_9 with funding from NextGenerationEU funds; the “Ministerio de Ciencia, Innovaci\'on y Universidades”  and its “Plan de Recuperaci\'on, Transformaci\'on y Resiliencia”; “Consejer\'ia de Universidad, Investigaci\'on e Innovaci\'on” of the regional government of Andaluc\'ia and “Consejo Superior de Investigaciones Cient\'ificas”, Grant CNS2023-144504 funded by MICIU/AEI/10.13039/501100011033 and by the European Union NextGenerationEU/PRTR,  the European Union's Recovery and Resilience Facility-Next Generation, in the framework of the General Invitation of the Spanish Government’s public business entity Red.es to participate in talent attraction and retention programmes within Investment 4 of Component 19 of the Recovery, Transformation and Resilience Plan; Junta de Andaluc\'{\i}a under Plan Complementario de I+D+I (Ref. AST22\_00001), Plan Andaluz de Investigaci\'on, Desarrollo e Innovación (Ref. FQM-322). ``Programa Operativo de Crecimiento Inteligente" FEDER 2014-2020 (Ref.~ESFRI-2017-IAC-12), Ministerio de Ciencia e Innovaci\'on, 15\% co-financed by Consejer\'ia de Econom\'ia, Industria, Comercio y Conocimiento del Gobierno de Canarias; the "CERCA" program and the grants 2021SGR00426 and 2021SGR00679, all funded by the Generalitat de Catalunya; and the European Union's NextGenerationEU (PRTR-C17.I1). This research used the computing and storage resources provided by the Port d’Informaci\'o Cient\'ifica (PIC) data center.
State Secretariat for Education, Research and Innovation (SERI) and Swiss National Science Foundation (SNSF), Switzerland;
The research leading to these results has received funding from the European Union's Seventh Framework Programme (FP7/2007-2013) under grant agreements No~262053 and No~317446;
This project is receiving funding from the European Union's Horizon 2020 research and innovation programs under agreement No~676134;
ESCAPE - The European Science Cluster of Astronomy \& Particle Physics ESFRI Research Infrastructures has received funding from the European Union’s Horizon 2020 research and innovation programme under Grant Agreement no. 824064.}

\end{document}